# Direct-Write Printed Contacts to Layered and 2D Materials

Sharadh Jois*, Erica Lee, Philip Li, Tsegereda Esatu, Jason Fleischer, Edwin Quinn, Genda Gu, Vadym Kulichenko, Luis Balicas, Son T. Le, Samuel W. LaGasse, Aubrey T. Hanbicki, Adam L. Friedman*

S. Jois, E. Lee, P. Li, T. Esatu, J. Fleischer, E. Quinn, S. T. Le, S. W. LaGasse, A. Hanbicki, A. Friedman
    Laboratory for Physical Sciences, 8050 Greenmead Drive, College Park, MD 20740, USA
        *Email:* sjois@lps.umd.edu, afriedman@lps.umd.edu

G. Gu
    Condensed Matter Physics and Materials Science Department, Brookhaven National Laboratory, Upton, New York 11973, USA
    Department of Physics and Astronomy, Stony Brook University, Stony Brook, New York 11794, USA

L. Balicas
    National High Magnetic Field Laboratory, Tallahassee, FL 32310, USA
    Department of Physics, Florida State University, Tallahassee, FL 32306, USA

V. Kulichenko
    National High Magnetic Field Laboratory, Tallahassee, FL 32310, USA

## Table of Contents

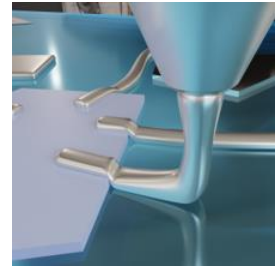

We demonstrate that direct-write printing is a viable method to fabricate devices on different 2D and layered materials yields Ohmic contacts in a single step, as compared to multi-step lithography processes which often yield poor contacts. Printed devices perform exceedingly well with gating, temperature cycling, and in magnetic fields. Our work motivates adoption of additive manufacturing for fabricating novel material devices especially for rapid material development and device prototyping.

## Abstract

Advancements in fabrication methods have shaped new computing device technologies. Among these methods, depositing electrical contacts to the channel material is fundamental to device characterization. Novel layered and two-dimensional (2D) materials are promising for next-generation computing electronic channel materials. Direct-write printing of conductive inks is introduced as a surprisingly effective, significantly faster, and cleaner method to contact different classes of layered materials, including graphene (semi-metal), $MoS_2$ (semiconductor), Bi-2212 (superconductor), and $Fe_5GeTe_2$ (metallic ferromagnet). Based on the electrical response, the quality of the printed contacts is comparable to what is achievable with resist-based lithography techniques. These devices are tested by sweeping gate voltage, temperature, and magnetic field to show that the materials remain pristine post-processing. This work demonstrates that direct-write printing is an agile method for prototyping and characterizing the electrical properties of novel layered materials.

# 1 Introduction

Novel two-dimensional (2D) and layered materials[1] are critical elements in future devices for advanced computing. Their inherent atomic thinness, varied properties, and discrete stackability can enable atomic-scale gate-all-around field effect transistors (FETs)[1,2], flexible electronics, new alternative state-variable computing paradigms[3–5], and emergent quantum properties amenable to low-power implementation[6,7] or advanced sensors[8,9]. Nonetheless, scaling the device fabrication to commercially viable levels is currently limited by sample size, uniformity, and reliable fabrication of electrically high-quality contacts to devices[10]. Work proceeds quite successfully on large-area growth methods, e.g. chemical vapor deposition (CVD)[11,12], yet reliable, simple contact formation remains elusive. Currently, there are several issues that still need to be solved despite ample research efforts.

First, although mechanically exfoliated flakes from bulk crystals are the highest-quality layered materials, positioning contacts on randomly scattered flakes is problematic for reproducibility and standardization of commercial lithography. Second, standard deposition techniques often yield non-Ohmic contacts. This could be due to damage induced through high energy or temperature patterning and deposition techniques common for device fabrication[13]. Studies have demonstrated defect formation during the deposition of metal films at elevated temperatures[14]. However, solutions to improve the quality of the contacts require additional cumbersome stacking steps[15–17] or using specific, industrially non-standard metals such as Sc[18], In[19] or Bi[20] to create better interfaces. Third, because 2D materials are entirely surface dominated and greatly affected by surface adsorbates[21,22], unavoidable resist residues can compromise device uniformity and behavior[23]. The common methods to create electrical contacts to layered materials all require polymeric resists for lithography, significantly exacerbating the adsorbate problem. Previous advances in lithographic contact engineering to 2D and layered materials demonstrated drastically improved quality of electrical contacts by eliminating polymers from contaminating the active channel material[14]. Glovebox fabrication,[24] hexagonal boron nitride[25] (h-BN) encapsulation and edge contacts[26], nano-via contacts[27], van der Waals (vdW) contacts[13,28], and work function engineering[18] have the desirable two-fold benefit of maintaining material cleanliness and improving contact quality. However, the stacking methods require many additional steps that consume several days of fabrication involving dry stamping, several steps of lithography to etch undesired areas, deposition of metal contacts, and lift-off. Moreover, device yields are exceptionally small. Each additional step adds failure modes, thwarting the rapid prototyping of new devices with layered materials.

The historic scaling of device dimensions starting from tens of microns in 1971 to the single nanometer regime today is indicative of the link between advancements in fabrication and the increase in device density[29]. Wafer-scale parallel patterning of devices primarily relies on several steps involving reticle lithography, etching, and depositions. Despite long processing times and strict parameters, the reward of creating arrays of high-density devices have made reticle lithography a scalable technique for high-volume manufacturing within the semiconductor industry. In contrast, sequential patterning techniques such as e-beam or direct-write optical lithography provide flexibility for prototyping custom devices, especially at the lab scale for testing novel materials. The emerging field of 2D and layered materials requires reliable techniques for making custom electrical contacts at the few-micrometer scale. Although the variety of layered materials has grown over the last decade, the fabrication methods of devices has largely been restricted to sequential lithography.

Additive manufacturing, or printing, is a sequential patterning method used in printed circuit boards and packaging of semiconductor chips, largely due to the high write speeds (1-100 mm/s) and sub-millimeter resolution. Advances in additive manufacturing capabilities[30] have offered scalable avenues to directly fabricate new electronic circuits and devices[31–33]. Printing devices can be a rapid route to fabricate devices on the large number of emerging novel materials. Previous reports[34–36] used ink-jet printing to make contacts and capacitors on $MoS_2$ films grown by CVD on solid and flexible substrates. Devices on CVD grown films on disordered substrates are not an ideal platform for testing basic device performance and material quality. Alternatively, direct-write aerosol-jet (AJ) printing[37] has emerged as another technique to print a variety of devices such as resistors, capacitors, sensors, and thin film transistors, on non-uniform surfaces. A variety of conductive and dielectric inks have been developed to meet these needs. AJ printed electronics are already integrated in high-frequency RF interconnects, chip packaging, and package shielding[32]. Testing direct-write AJ printing of microscale contacts to fabricate devices on high-quality 2D materials has remained unexplored.

Here, we will investigate AJ direct-write printing to align and deposit high quality contacts on exfoliated flakes of 2D materials with speed, accuracy, and reproducibility. This technique reduces the process of making microscale electrical contacts to a single step. An advantage of AJ printing is the versatility of using different metal and alloy nanoparticles[38,39] in a solvent that favors aerosolization, allowing us to customize the properties of the ink for contact engineering. In this study we use AJ direct-write printing[40] of silver nanoparticle (AgNP, diameter 40 nm) inks to create electrical contacts to flakes of different layered and 2D materials displaying distinct electronic behavior to demonstrate the broad applicability of the method. The materials tested in the study include graphene (semi-metal), $MoS_2$ (semiconductor), $Bi_2Sr_2CaCu_2O_{8+x}$ (Bi-2212, BSCCO, high-$T_c$ superconductor), and $Fe_5GeTe_2$ (FGT, metallic ferromagnet). The quality of contacts is assessed by two-terminal DC I-V curves for all devices, and the contact resistance for each material is compared with standard methods. Further electrical characterization on printed devices also shows their performance is comparable to traditional direct contacts. Specifically, we tested the electrostatic gate response of graphene and $MoS_2$ devices, measurement of the superconducting transition in BSCCO, and the anomalous Hall effect (AHE) in FGT. The cryogenic measurements on BSCCO and FGT, furthermore, validates printed contacts are appropriate at the low temperatures required for condensed matter experiments. Through these results, we demonstrate that printing contacts is a reliable approach to making high-quality devices on various novel materials, without the limitations of standard lithography such as fixed mask design, resist residue, and high-temperature metal deposition.

## 2 Results and Discussion

### 2.1 Device Fabrication through Lithography as compared to Printing

The steps for the fabrication of contacts to exfoliated materials using standard lithography as compared to direct-write printing are depicted in the flowcharts in **Figure 1 (a) and (b)**, respectively. Common to both routes are obtaining flakes of layered materials and identifying candidates for device fabrication. We begin with an adhesive tape-based method to exfoliate all chosen materials on $SiO_2$ wafers with pre-defined alignment features. Additional details are given in the methods section. Designs for contacts to candidate flakes are then created, with appropriate consideration for critical dimensions.

The procedure for standard lithography is outlined in **Figure 1(a)** and includes five steps. Below is the basic procedure and some issues relevant for 2D and layered materials:

1. Spin-on Photoresist: This step involves spin-coating the flakes with photoresist and baking off the solvent. Some 2D materials cannot survive at the required baking temperatures.
2. Expose: To establish the desired arrangement of contacts, the photoresist must be exposed to UV light or electron beams, usually with a reticle or photomask with fixed device designs, or with more advanced direct-write lithography tools.
3. Develop: The desired contact pattern is revealed by dissolving the exposed photoresist in a solvent. The residual photoresist on the surface of the active material can add significant contact resistance. While working with bulk materials in standard lithography, the residue is often removed by oxygen or argon plasma, adding a step in the process after (3) and before (4). Plasma exposure of ultra-thin layered materials can cause severe damage to the pristine surfaces and yield poor-quality devices. Thus, dose and development conditions must be precisely optimized to arrive at specific recipes that yield good devices on each separate material. Another key aspect of the development procedure is to ensure that there is an undercut in the sidewalls of the exposed resist to allow for the subsequent metal film to lift-off.
4. Metal Evaporation: Deposition of metallic films form the contacts in the pattern prescribed by the established polymer mask. Most common metal deposition methods are performed in high vacuum at high temperatures, and with a high energy source. Hot, vaporized metal can induce further damage to the exposed surface of ultra-thin materials[41], leading to trapped states, Fermi-level pinning[42,43], and randomness in efforts for Schottky-barrier engineering in semiconducting 2D materials. Pristine metal films such as Au and Pt, also require Ti or Cr as precursor layers to promote adhesion to the substrate. Often, these adhesion layers are not ideal for good electrical contacts. Other metals, like Ag which makes better contact with BSCCO[44], need to be deposited without an adhesion layer and a second lithography step is required to make the Ti/Au contact pads to connect probes or wires. Newer deposition techniques can combat problems of poor contact resulting from aggressive deposition methods. These include transferring layered materials onto pre-fabricated metal contacts[13,28] or stamping h-BN embedded with metal-via contacts[27] onto select material, routes that depend on another non-scalable technique. Other alternatives include using graphene or other metallic 2D contacts[17] or thermal nanolithography[45].

5. <u>Lift-off</u>: The metal-photoresist bilayers are stripped from unexposed areas using solvents, leaving behind the target flake with the desired metallic contacts. Care must be taken to avoid accidentally delaminating the exfoliated flakes during this step. Common methods to enhance lift-off such as ultrasonic cleaning, snow jet cleaning, or wiping are not possible.

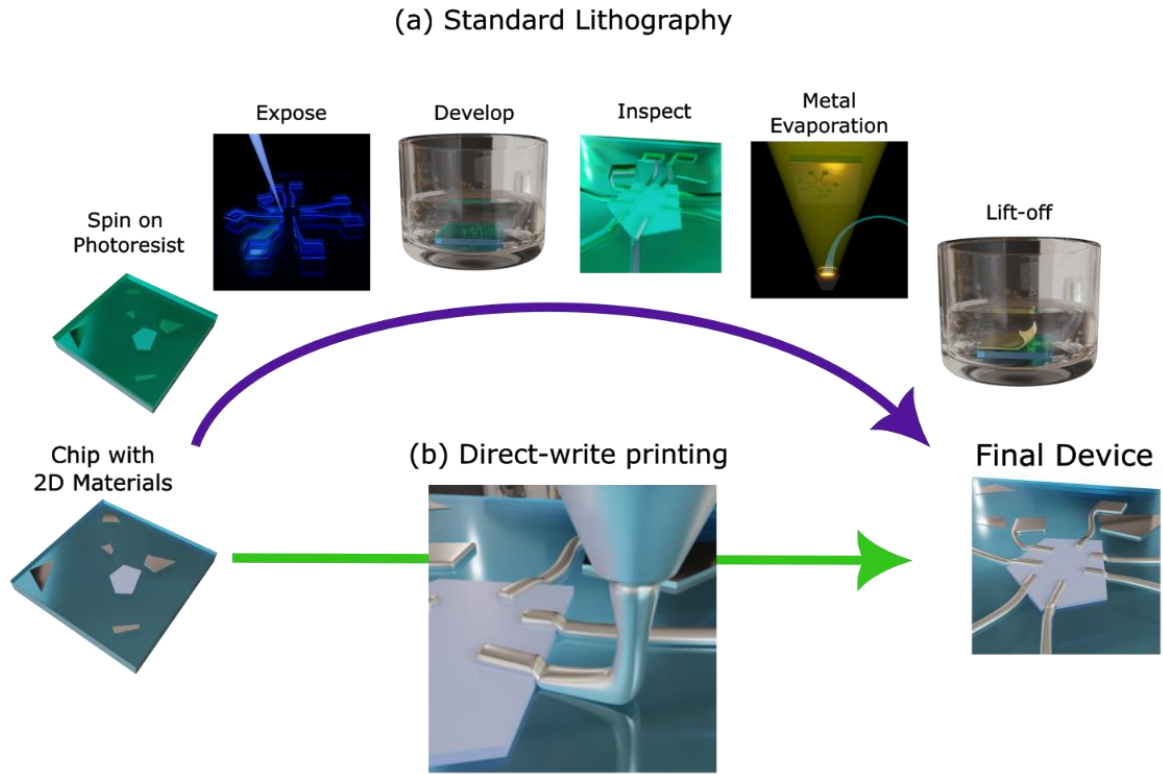

Figure 1. Comparison of workflow between (a) direct-write lithography and (b) direct-write printing to fabricate contacts to 2D materials on a chip. In the lithography route, polymeric photoresist is spin-coated on to the sample. The sample is then exposed to high energy beam of UV light or electrons which rasters over the target area to transfer the device design. The exposed regions of the resist are then dissolved in a developer solvent. This creates openings in the resist where metallic films for contacts are deposited, usually by high-energy electron beam evaporation in an ultra-high vacuum chamber. Finally, the sample is immersed in a solvent to lift-off the metal and resist film stack and leave behind metal contacts on finished devices. Contrarily, using direct-write printing contacts can be deterministically deposited to create devices in a single step.

In contrast to standard lithography, direct-write printing has a single step depicted in **Figure 1(b)**. The process consists of a printer that directs an ink stream to print each feature sequentially, removing stitching errors common in raster scan direct-write techniques and rendering on-demand flexibility to device designs. Although there are several direct-write printing methods, such as ink-jet printing and extrusion-based syringe printing, we chose AJ printing because it has a large working distance (1-5 mm) and ability to print a minimum linewidth of $d = 20$ µm. Notably, submicron resolution is possible with more advanced techniques such as electrohydrodynamic jet printing [32,46,47] or capillary flow printing.

One of the concerns with AJ printing is the region of ink overspray around the intended print line, systematically studied by our group[48]. We performed careful optimization at the beginning of each write to minimize overspray, which can be $l_{os}$ ≤5 µm in best cases or be sparsely distributed up to tens of microns in other cases. To combat losing devices to overspray, we strategically designed the devices to space the contacts such that the device length $L \geq (l_{os} + d + l_{os})$ and the flake length is $L_{flake} \geq (L + 2d)$. For our printer, a minimum separation $L \sim 30$ µm between any pair of contacts was sufficient to remove any risk of shortening through the overspray region. The selected flakes were at least 50 µm in length along one dimension. Optical inspection is used to select devices that show a clear channel region suitable for experiments. The contacts shorted by overspray on the channel of layered materials are highly conductive and do not permit a systematic study of short channel effects. Thus, they are disregarded in our experiments.

The AJ printing technique is as simple as first priming with ink to minimize deviations in deposition rate, adjusting the process gas flow rates to achieve the desired linewidth and minimum overspray, and then aligning the substrate and printing the devices. The AJ printer used three known coordinates of pre-patterned markers on the samples, similar to direct-write lithography, for accurate device alignment. Once printing is finished, the inks are annealed at 150 °C for 1 to 3 hours in vacuum to remove the solvent. The annealing conditions used in this paper are specific to the solvent (butyl carbitol) used for the ink and the material's sensitivity to degradation. See Methods section 2.8 for more details. There are a variety of silver inks available that have milder annealing requirements and can be used for applications that involve polymer-based substrates where the temperature limitation is lower. Some examples of flexible substrates used in printed electronics are polyethylene terephthalate (PET), polyethylene naphthalate (PEN), and polyimide (PI). The temperature limitation for these substrates based on the glass transition temperature are 100 ºC, 120 ºC, and 200 ºC, respectively. In these applications, an ink with milder annealing conditions can be selected. For example, Electroninks has an array of silver inks that can be cured at 100 ºC. Alternatively, other water-based inks using novel nanomaterials could be explored for a more gentle processing [49]. The methods section contains detailed information on controlling the linewidth, ink quality, and curing steps. Overall, the multi-step process and challenges faced during standard lithography, such as resist residue and high-temperature metal deposition, are avoided by using this low energy, direct-write AJ printing.

To validate and benchmark the direct-write approach, we assessed the quality of printed contacts to different materials by measuring the two-terminal current-voltage (*I-V*) behavior and determining the contact resistance for each material. The length dependence on contact spacing can be inferred from supplemental Figure S1, where the slope provides the sheet resistance and the intercept the contact resistance. Any pair of shorted contacts were eliminated from our experimentation. We also made material-dependent measurements with more elaborate devices to elucidate the generality of this approach. Specifically, we performed gate-dependent transport measurements on monolayer graphene and thin $MoS_2$, determined the superconducting transition in BSCCO with a four-point measurement of resistance as a function of temperature, and probed the anomalous Hall behavior in FGT as a function of the magnetic field on a 6-terminal device. In all these measurements, the printed contacts performed admirably and were robust to a variety of conditions such as cryogenic temperature cycling and high magnetic field.

## 2.2 Semi-metal – Graphene Device

Graphene is a single layer of carbon atoms covalently bonded in a hexagonal lattice. It can be isolated from bulk graphite by tape-based mechanical cleaving, and details of this process are widely available in the literature[50]. The electrical properties of graphene, a gapless Dirac semi-metal, are also widely studied with vast amounts of details available elsewhere in the literature[51,52]. One-dimensional edge-contacts[26] to encapsulated graphene has become the state-of-the-art method for making high-quality graphene devices in the lab. However, this method is not yet scalable. The creation of good quality devices on graphene by AJ printing introduces an alternative to current methods. Additionally, it can also allow the creation of graphene devices on non-flat and flexible substrates for applications in sensing[8,9,53].

An optical micrograph of a printed graphene device on a 90 nm $SiO_2$/Si n++ substrate is shown in **Figure 2(a)**. The Ag-based ink has printed linewidths of ~20 um separated by $L$~27 um. The circular features at the ends of the contacts are bumps formed due to the longer dwell time of the AJ ink stream before the shutter closes. The Ohmic quality of the contacts to monolayer graphene is confirmed by the representative linear I-V curve in **Figure 2(b)** where DC bias voltage was swept across two contacts, taken at room temperature. Making this measurement on a series of samples with different lengths, L, we estimate that the contact resistance to graphene is 1.88 kΩ µm. A more complete discussion of the contact resistance measurements is made below and in the Supplement.

The devices are fabricated on a layer of $SiO_2$ with a highly doped Si substrate. Therefore, we can apply a global backgate capable of introducing carrier densities up to $n \sim 10^{13}$ cm$^{-2}$. We measure the device in a vacuum probe station after annealing at 400 K for 2 hours to desorb any atmospheric gases on the graphene film. When we sweep the back-gate voltage ($V_{bg}$) from negative (hole doping) to positive (electron doping) and back (**Figure 2(c)-inset**) while maintaining a constant source-drain bias of 1 mV, we measure a symmetric resistance as shown in **Figure 2(c)**. The gate voltage at resistance maximum corresponding to the Dirac point ($V_{dp}$) and has been adjusted on the horizontal axis. The position of the Dirac point at $V_{dp}$ = -3.5 V indicates the processed graphene device is minimially hole-doped, which is commonly the result of charge traps in the $SiO_2$ dielectric. Such deviations of the Dirac point are also seen in lithographic devices of bare graphene on $SiO_2$. In fact, for most graphene devices in the literature, excessive charge traps on the substrate (an extra problem for electron beam lithography methods) and residual photoresist from standard lithography result in significant hysteresis in the transfer curve.

Our device's electrical response is in excellent agreement with the charge impurity scattering model[54,55] that explains the transport behavior of graphene on $SiO_2$ substrate. The conductivity σ($V_{bg}$) at high carrier density could be fitted with a standard model[55] where $\mu_e$, $\mu_h$ are the electron and hole field-effect mobilities, $c_g$ is the geometric gate capacitance per unit area, 3.84 x 10$^{-4}$ Fm$^{-2}$, σ$_{res}$ is the residual conductivity that is determined by the fit. Our device has $\mu_e = 5240 \pm 15$ cm$^2$V$^{-1}$s$^{-1}$ and $\mu_h = 6300 \pm 30$ cm$^2$V$^{-1}$s$^{-1}$ and interestingly the ratio $\frac{\mu_e}{\mu_h} \sim 0.83$ is in good agreement with high quality graphene on $SiO_2$ devices reported in literature[55,56]. Here, by using a process free of photoresist, we observe neglibible hysteresis and significantly less charge trapping. This confirms that printed graphene devices can be dynamically gated into the valence or conduction bands. Our results, reproduced with multiple devices, provide evidence that good quality gate-tunable graphene devices can be realized with direct-write printing. Refer to the supplemental information for results highlighting the durability of printed graphene devices.

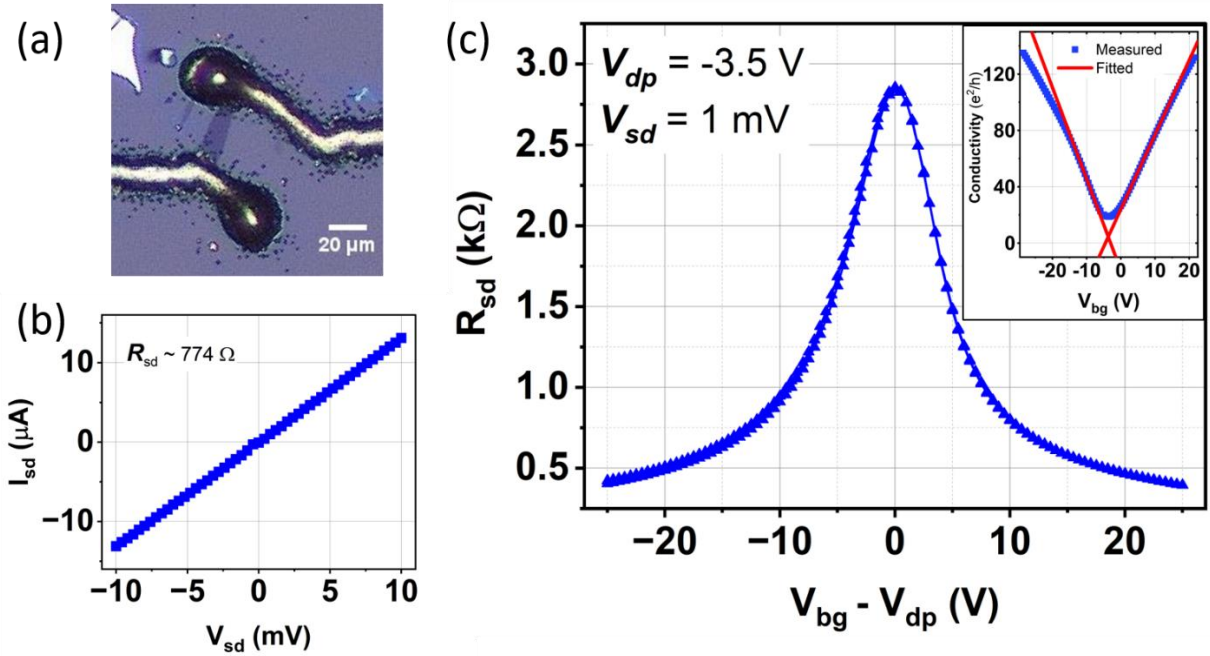

Figure 2. (a) An example of a printed graphene device. (b) Two-terminal I-V curve to show the Ohmic quality of the contacts to graphene. (c) Room-temperature transfer curve with negligible hysteresis on graphene showing ambipolar response to sweeping the back-gate voltage ($V_{bg}$) taken at constant source-drain ($V_{sd}$) bias. The horizontal axis has been centered at the Dirac point ($V_{dp}$). The inset shows corresponding the conductivity and fit curves (red) for device mobility.

## 2.3 Semiconductor - MoS$_2$ Device

2D semiconductor materials of the transition metal dichalcogenide (TMD) family, such as MoS$_2$, WS$_2$, and WSe$_2$, are promising for beyond-silicon computing applications. Particularly, their atomic thickness and being free of dangling bonds can reduce short-channel effects[2] and their mobility is maintained even to atomic scales[57]. However, achieving high-quality contacts to 2D semiconductors, free of Fermi-level pinning and large Schottky barriers due to metal-induced gap states[42], is an ongoing challenge[14]. One approach is to use van der Waals contacts like graphene[17] or transferred metal films[19,58,59]. Other solutions engineered to combat these issues have semi-metal[20] contacts or introduced a removable buffer layer[28,60]. All these methods suffer from great complexity. However, simple AJ printing of Ag conductive ink to a TMD semiconductors results in Ohmic contacts.

Like the graphene, A two-terminal device on a semiconducting flake of thin (7.4 nm) $MoS_2$ (measure thickness) on 90 nm $SiO_2$, shown in **Figure 3(a)**, is tested for gating behavior in a high vacuum chamber. The initial gating tests performed in ambient atmosphere at different source-drain voltages emphasizes the high quality of the printed contacts. The sample was annealed in-situ at 400 K for 1 hour to remove atmospheric adsorbates which can create dipoles and contribute to hysteresis. The transfer curve resembling a conventional field-effect transistor (FET) device is obtained with a switching current ON/OFF ratio greater than $10^6$, (**Figure 3(b)**) taken with a small source drain voltage $V_{sd} = +0.5$ V. Supplemental Figure S3 (a) shows similar high ON/OFF ratio curves taken during initial tests in ambient atmosphere with $V_{SD} = -0.5\ to +0.5$ V. The improvement in device behavior can be seen in Supplemental Figure S3 (b). The inset shows the same data on a linear scale. Due to the large bandgap, $E_g \sim 1.9$ eV, and native n-doping in $MoS_2$, gating into the hole branch is not observed even at $V_{bg} < -50$ V. The saturation region of the transfer curve is fit to the n-type 2D FET equation $I_{sd} = \frac{\mu_n W C_{ox}}{L}[(V_{bg} - V_{th})V_{sd}]$ to extract the mobility $\mu_n$ and threshold voltage $V_{th}$. In the forward sweep going from $-50$ V to 0 V, $\mu_n = (33.78 \pm 0.02)$ cm$^2$/Vs and $V_{th} = -40.6$ V, based on the black curve. The fit for the reverse sweep 0 V to $-50$ V, yields $\mu_n = (35.27 \pm 0.02)$ cm$^2$/Vs and $V_{th} = -38.4$ V, based on the red curve.

The two-terminal *I-V* curves from this device shows nearly Ohmic behavior at several gate voltages, **Figure 3(c)**, including near the OFF state, **Figure 3(d)**. The Ohmic nature of the contacts and the high switching ratio with gating verify we can easily create high quality FET devices to 2D semiconductors using printed contacts. These results show the mobility of printed devices on thin $MoS_2$ is comparable to previous results of high quality devices made by lithography on $SiO_2$ substrates[61]. Additionally, in previous reports [34–36] of $MoS_2$ devices created by inkjet printing on $SiO_2$ or on paper, $\mu_n < 5$ cm$^2$/Vs and ON/OFF ratio was ~ $10^4$ and required extremely large $V_{sd}$ ranging between 1 V to 20 V. The AJ printed $MoS_2$ device discussed here shows higher metrics in all aspects.

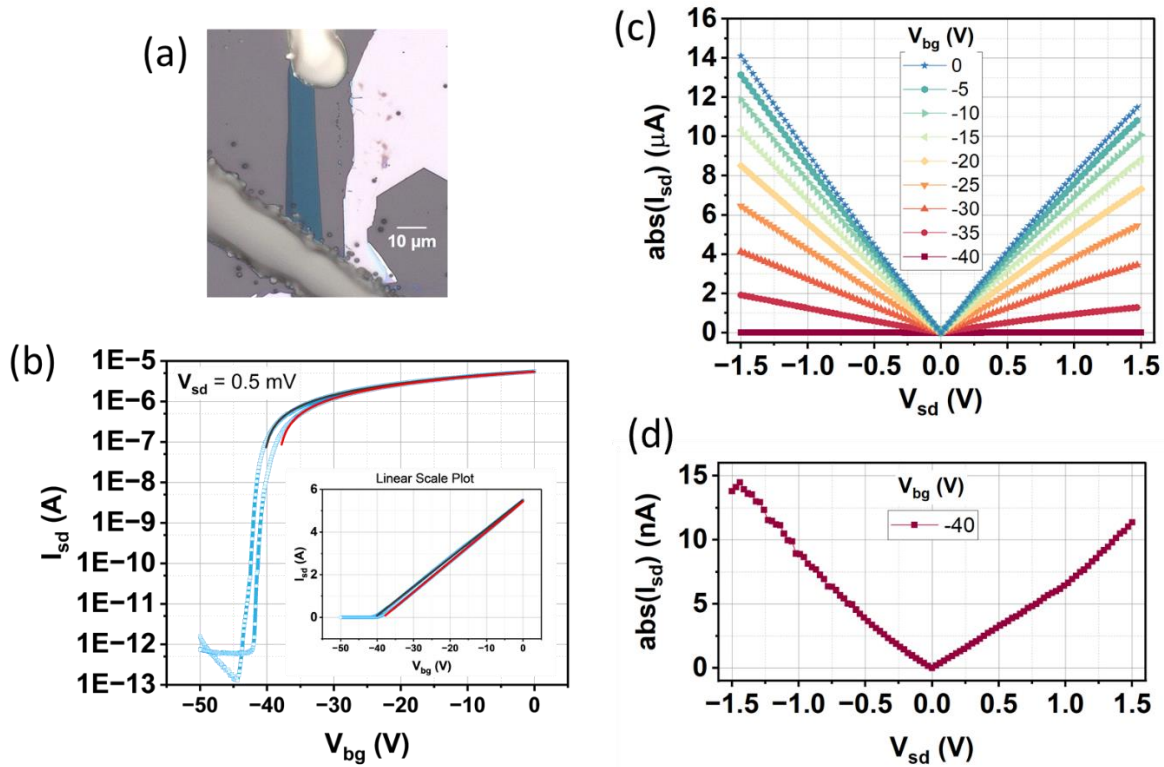

Figure 3. (a) Thin (7.4 nm) MoS$_2$ device with printed contacts. (b) Transfer curve corresponding to the thin MoS$_2$ device shown in (a) having $\frac{I_{on}}{I_{off}} > 10^6$ and heavy n-doping. The inset shows the same data in a linear plot. The Ohmic nature of the contacts is confirmed by the linear response in current to sweeping source-drain voltage at different back-gate voltages ($V_{bg}$) in (c). The data at $V_{bg} = -40$ V are rescaled separately in (d).

## 2.4 Superconductor – BSCCO Device

The development of exfoliation and stacking techniques for graphene and TMDs has also led to advancements in layered superconductors such as BSCCO and NbSe$_2$. Like graphene, these are also van der Waals materials and can be exfoliated using adhesive tape. Layered superconductors are promising for next-generation quantum devices and sensors because of atomically pristine surfaces and edges, and higher critical fields and transition temperatures compared to conventional deposited superconducting films such as Al and Nb. BSCCO is a high-T$_c$ cuprate nodal superconductor[62] whose superconducting properties are strongly dependent on the oxygen concentration[63]. Superconductor-to-insulator transition by loss of oxygen from BSCCO[64] provides convincing evidence that extreme precautions are necessary to preserve its superconducting properties. An inert environment, such as an Ar-filled glovebox with low H$_2$O vapor (like our set-up) or UHV chamber, is needed to handle BSCCO during exfoliation. Furthermore, creating

Ohmic contacts to BSCCO requires special steps such as removing degraded layers by wet-etch[65] or in-situ Ar milling before metal deposition[66], stacking on pre-fabricated gold electrodes at cryogenic temperatures[67], or creating microscale hard masks and depositing metal at cryogenic temperatures[68]. We demonstrate that the rapid and gentle nature of AJ printing of Ag contacts can reliably achieve ohmic contacts to exfoliated BSCCO flakes while preserving the superconducting properties, without the need for additional fabrication.

We obtain large areas of exfoliated BSCCO flakes inside an Argon-glovebox and fabricate 4-terminal devices, such as the one in **Figure 4(a)** of thickness 414.8 nm. I-V curves for three contact combinations in **Figure 4(b)** show they are Ohmic and capable of injecting high currents. We then observe the superconducting transition as expected at $T_c \sim 90$ K, shown in **Figure 4(c)**. We repeated this measurement several times, and the printed contacts maintained their stability through at least three iterations of thermal cycling.

Although our results are on bulk BSCCO flakes, we recognize that thin layers of BSCCO can be instantly depleted of oxygen in ambient conditions or at high temperatures. The annealling conditions used here are compatible with bulk flakes, see methods section. Further process development is required to correlate annealing conditions to the thickness of BSCCO while tracking its $T_c$. Ultimately, exploring direct-write printing while holding the sample at cryogenic temperatures in a vacuum chamber may be required to work with few layers of BSCCO. Additionally, the solvent would have to evaporate at cryogenic temperatures. Nonetheless, our results on AJ printing AgNP contacts to thicker BSCCO provides a reliable way to fabricate devices based on a high-$T_c$ superconductor. See the supplemental information for contact resistance and reproducibility of printed devices on bulk BSCCO. Other popular layered superconductors, $NbSe_2$ and $Fe_3Te_{0.45}Se_{0.55}$ for example, that are less sensitive to degradation may also yield good devices through AJ printing.

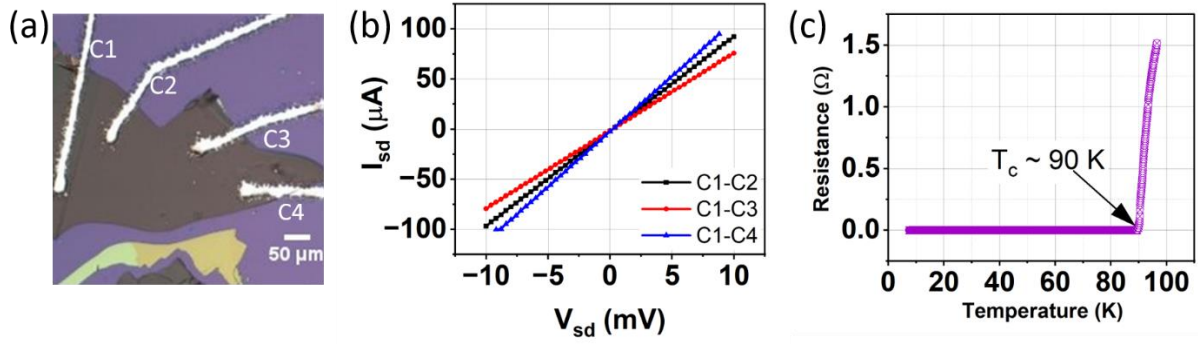

Figure 4. Examples of printed device on bulk (414.8 nm) BSCCO flake. (b) Two-terminal I-V curves to show the Ohmic quality of the contacts to device in (a). (c) Superconducting transition observed in printed BSCCO device at 90 K.

## 2.5  Ferromagnet – FGT Device

Magnetic 2D materials are promising for novel spintronic logic and memory devices[69] that are faster and more power efficient compared to current computing architectures[5]. $Fe_3GeTe_2$, $CrSBr$, $CrI_3$, $CrGeTe_3$, and more recently $Fe_5GeTe_2$, are all emerging layered magnetic crystals that can be isolated by mechanical exfoliation. Some of these materials are prone to rapid degradation when exposed to air during long fabrication steps in standard lithography. The development of a reliable and scalable process to make contacts to flakes and films of magnetic 2D materials will be invaluable for fundamental studies and next-generation spintronic devices.

$Fe_5GeTe_2$ (FGT) is fairly air-stable and undergoes a ferromagnetic (FM) transition near room temperature[70–72]. Multiple contacts were printed to create the six-terminal device shown in **Figure 5(a)**. The DC I-V curves measured between C1 and other contacts in **Figure 5(b)** confirm they are all Ohmic. To confirm magnetization behavior, we measure the Hall resistance of the device, **Figure 5(c)**, and observe the anomalous Hall effect, AHE[73]. In this measurement, an in-plane current is applied to the film while the transverse ($V_{xy}$) voltage is monitored in an externally applied, sweeping, perpendicular magnetic field, $H_z$. For a ferromagnetic film, spin-dependent scattering and/or quantum mixing alter the Berry phase curvature of the material as compared to a film with no internal magnetic field, resulting empirically in an additional term in the usual Hall voltage relation: $\rho_{xy} = R_H H_z + R_s M_z$. Here, $\rho_{xy}$ is the Hall resistivity, $R_H$ is the Hall coefficient, $R_s$ is a material specific proportionality coefficient, and $M_z$ is the internal magnetization created by the relativistic spin-orbit interaction[73]. This second term is the AHE and phenomenologically, the resulting curve is an exact reproduction of the hysteresis loops attained through standard magnetometry. The advantage of this electrical measurement method is it allows the measurement of signals that are too small to be measured using other magnetometry methods, because the signal from a 2D film is very small due to its inherently low volume. In our measurement, the absence of coercive field and increasing saturation field from 0.3 T at 250 K to 0.7 T at 100 K, agree with previous reports[72,74] on bulk FGT. This result demonstrates that printed contacts work well on magnetic layered materials, and the contacts are robust to thermal cycling in an applied magnetic field.

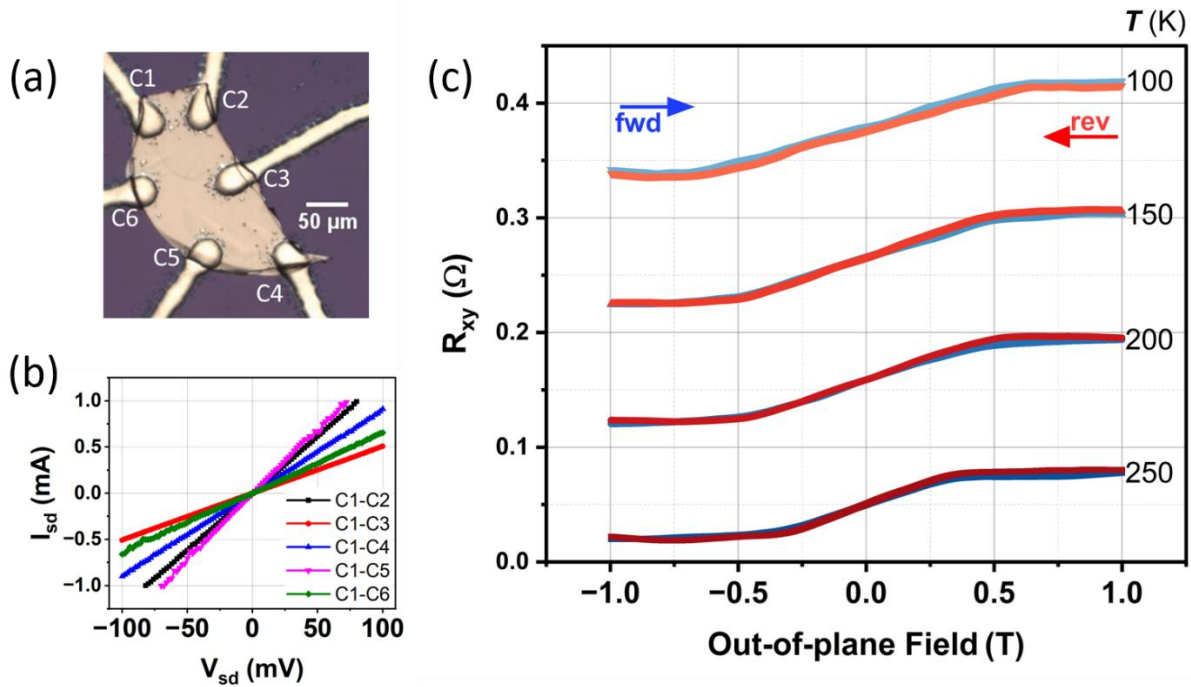

Figure 5. (a) Six-terminal device on large bulk FGT flake. (b) Two-terminal I-V curves to show the Ohmic quality of the contacts to device in (a). (c) The hall resistance ($R_{xy}$) measured on the multi-terminal device in (a) taken at different temperatures and offset by 0.1 Ω for clarity. The forward (fwd) and reverse (rev) sweeps of the magnetic field are in blue and red respectively.

## 2.6 Contact Resistance Analysis

For all the samples, we employ a simple approach to estimate the contact resistance, $R_c$. Using the source-drain resistance ($R_{sd}$) measured between two contacts separated by length, $L$, the linear relation $R_{sd} \times W_{sd} = 2R_c + R_{sheet} \times L$ can be used to fit the experimental data. The derivation of this relationship can be found in the supplemental information. Here, $W_{sd}$ is the average width of the arbitrarily shaped flake between the source-drain contacts, and the slope $R_{sheet}$ is the sheet resistance that agrees with 4-wire measurements. Although the channel resistance is a function of $L$, the constant term $R_c$ is a reasonable estimate for contact resistance. Due to the large linewidth of the printed contacts, we first tested the feasibility of our technique on large-area bulk graphite flakes that are abundant on exfoliated samples and measured $R_c = 196.6 \pm 11.7$ Ω μm (See Supplement). The low contact resistance of graphite flakes suggests this method should work well on other materials and thinner flakes. Several devices with two or more contacts were printed on different materials for further analysis. $R_c$ found in printed contacts based on analysis in **supplemental Figure-S1** is compared with direct contacts in **Figure 6**. The direct contact method refers to standard lithography and deposition.

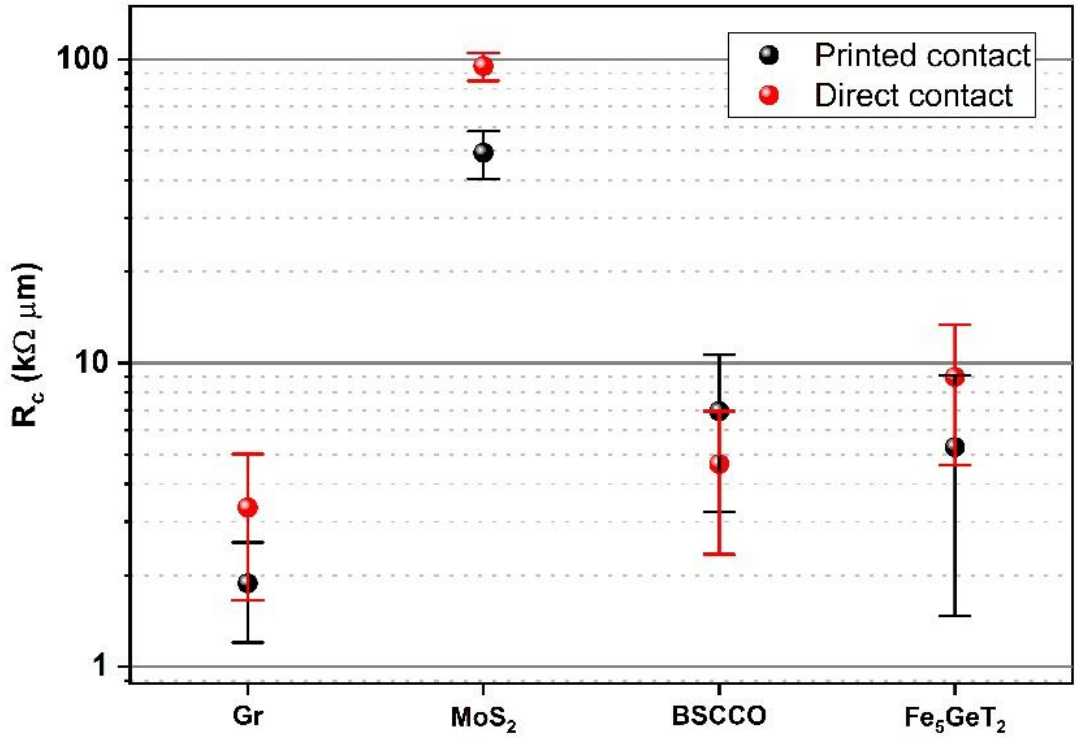

Figure 6. A scatter plot comparing the contact resistance to graphene (Gr), MoS$_2$, BSCCO, and FGT. The values of R$_c$ for direct contacts were obtained from literature for Gr, MoS$_2$, and BSCCO and citations are provided in the main text.

The estimated $R_{c,Gr} = 1.9 \pm 0.7$ kΩ μm for printed contacts to graphene is 56% of the contact resistance compared to Au direct contacts with Ti or Cr adhesion layers[27]. Similarly, we find $R_{c,MoS_2} = 49.2 \pm 8.8$ kΩ μm is also approximately half of what is reported for Ti/Au direct contacts to MoS$_2$[10,75]. Indeed, the gating results for these two materials, as discussed earlier, are made possible due to the high quality of printed contacts. The contact resistance observed on bulk BSCCO flakes is $R_{c,BSCCO} = 6.9 \pm 3.7$ kΩ μm, which is on par with results from more complex methods discussed earlier[65,76] also using evaporated Ag and Au bilayer contacts. For our contacts printed on exfoliated bulk FGT flakes, we achieve high-quality Ohmic contacts, with a $R_{c,FGT} = 5.3 \pm 3.8$ kΩ μm, also 40% less compared to direct contacts ($8.9 \pm 4.3$ kΩ μm) realized using evaporated Ti/Au in control devices.

## 2.7 Conclusion

We have demonstrated that AJ printing Ag ink on several sensitive low-dimensional materials yields Ohmic contacts. We confirmed the quality of materials remain pristine after fabrication through standard characterization. Ambipolar gating for graphene through the Dirac point was observed down to cryogenic temperatures. Multi-terminal printed contacts on high quality graphene should allow measurements of quantum Hall effect and other exotic properties. An ON/OFF ratio of $10^6$ was observed while gating a printed $MoS_2$ device, while also retaining Ohmic contact behavior at all gate voltages. This implies that printing FET devices on 2D materials for logic and sensors is a viable option. Furthermore, we also demonstrated that printing ohmic contacts to degradation sensitive layered materials such as BSCCO and FGT is also possible. Superconducting transition in BSCCO is seen at $T_c \sim 90$ K and AHE response of FGT is observed between 100-250 K where peak magnetization occurs. Through all of these measurements, the printed contacts remained robust through thermal cycling and even in a magnetic field.

With the recent advent of electrohydrodynamic (EHD) jet [46] and capillary flow[77] printing technology, we predict printing contacts with sub-micron resolution on layered materials will soon be possible. Additionally, metallic, polymeric, and dielectric inks are also available for AJ and EHD jet printing. Using actuated multi-nozzle print heads, it may soon be possible to print conductive and dielectric inks to print a field-effect transistor on semiconducting 2D materials. Multi-nozzle print heads where each ink-stream is independently actuated can help evolve AJ printing towards parallel patterning of multiple devices simultaneously. With the ever-growing menu of low-dimensional materials, opting to print contacts could significantly ease device fabrication for rapid prototyping and also allow the adoption of non-flat and flexible substrates in applications for sensors. AJ printing can be a solution to add contacts on 2D materials in scanning probe experiments where keeping the surface pristine after resist-based lithography is challenging.

## 2.8 Methods

The bulk layered crystals are cleaved and spread out using adhesive tape. Then flakes are exfoliated on plasma-treated 90 nm or 285 nm $SiO_2$ substrates with predefined alignment marks. BSCCO and FGT samples are prepared in an argon glovebox. These remain inside until the printer is ready to minimize oxidation risk. Flakes are identified using a microscope inside the glovebox. Graphene and $MoS_2$ samples are exfoliated in ambient and stored in a nitrogen-filled dry box. Flakes are chosen based on thickness, size, and separation from neighbors to meet patterning constraints and device requirements. Microscope images of the flakes are taken such that one or more alignment features are present in the field of view. These images are imported into CAD software and overlaid with the design file using the alignment features. The contacts to the chosen flakes are drawn with commercial CAD software.

We used commercial OPTOMEC HD2x and AJ5x aerosol-jet printers during this work. The process begins by aerosolizing a solvent-based AgNP ink held in a reservoir at room temperature using pneumatic or ultrasonic treatment. The aerosolized ink is transported through a mist tube by a carrier gas. We take several measurements to minimize the linewidth by optimizing the working distance, nozzle design, and temperatures of both ink and substrate. The aerosolized droplets are carried through a series of multi-stage focusing apertures and finally forced out from a nozzle through a 150 μm aperture. The working distance of the nozzle is between 2 to 5 mm above the substrate. The AgNPs with average diameter of 40 nm are aerosolized and deposited as droplets with diameters raging between 0.5 and 1 μm on the substrate. The AgNPs are suspended in butyl carbitol (72:28) due to its low volatility, good coalescing properties, and high solubility. The nominal values used for the sheath, atomizer, and exhaust flow rates were 30, 1200, 1150 sccm respectively. The speed of the sample platen during printing is between 1 to 3 mm/s.

Other geometries of silver nanomaterials were proposed and tested as part of this work and prior work[78]. One example is polymer ink (e.g. PEDOT) embedded with silver nanowires. The downside of silver nanowires is the rapid accumulation of aerosolized ink in the mist tube as well as at the mouth of the nozzle, leading to higher chances of clogging. AgNP inks were the most stable and easy to control in our laboratory's printers.

Once the printer is ready and the test print of the design is approved, we print on substrates carrying the layered materials. BSCCO and FGT samples are taken out of the glovebox and brought to the printer at this stage. The sample is held at 50 °C during the printing process to promote adhesion. Two or more passes are made over the probe pads to make them robust. We have optimized the conditions of the printer to achieve line widths of ≈20 um. Lastly, the contacts are annealed at 150 °C for 1 to 3 hours to cure the ink before electrical testing. Graphene, MoS$_2$, and FGT devices are annealed in a vacuum (1E-3 mbar) whereas BSCCO devices are annealed in an oxygen-rich atmosphere (1.5 to 2 mbar). The resistivity of the AgNP ink has been measured and calculated by the authors using 2-point (dog-bone style) and 4-point probe geometries. The resistivity of the printed ink is approximately 4X resistivity of bulk Ag. This result has been repeatedly demonstrated across a variety of substrates and agrees with the data provided by the manufacturer. The resistivity of the AgNP ink used here is 48 mΩ μm, negligible in comparison to the contact resistance of devices reported above.

In this study, we fabricated several devices of each type for process development and reproducibility. The devices are first tested in a room-temperature probe station for contact resistance. Optimal devices are loaded in cryostats for temperature and magnetic field-dependent characterization. Gold wires capped with indium solder are bonded onto the printed pads to connect the devices to terminals on a chip carrier. The ink is strongly bonded to the SiO$_2$ substrate without an adhesion layer and can withstand probing, pressing, and even wire-bonding.

## Supporting Information

Supporting Information is available from the Wiley Online Library or from the author.

## Author Contributions

S.J., A.L.F., and E.Q., conceived the idea. S.J. designed the experiments and collected the electrical data with assistance from T.E. Printing contacts on devices was done by E.L, P.L, J.F. S.W.L prepared the animated figure panels. S.T.L. fit the graphene transport data. V.K. and L.B. provided the FGT crystal. G.G provided the BSCCO crystal. S.J., A.T.H., and A.L.F prepared the manuscript.

## Acknowledgements

The authors acknowledge Konnor T. Kim, Gregory M. Stephen, and Vinay Sharma for discussions and assistance during the course of this work. Work at Brookhaven is supported by the

<seg data-type="">Office of Basic Energy Sciences, Materials Sciences and Engineering Division, U.S. Department of Energy (DOE) under Contract No. DE-SC0012704. L.B. acknowledges support from the US DoE, BES program through award DE-SC0002613 US (synthesis and measurements), US-NSF-DMR 2219003 (heterostructure fabrication), and the Office Naval Research DURIP Grant 11997003 (stacking under inert conditions). The National High Magnetic Field Laboratory acknowledges support from the US-NSF Cooperative agreement grant number DMR-2128556 and the state of Florida.</seg>

## Conflict of Interest Statement

The authors declare no conflict of interest.

## Ethical Statement

We declare that the ideas and data presented in this article is original work of the authors.